# EMPIRICAL PROBABILITY MODEL OF THE COLD PLASMA ENVIRONMENT IN JOVIAN INNER MAGNETOSPHERE

Yoshifumi Futaana [1], Xiao-Dong Wang [1], Elias Roussos [2], Pete Truscott [3],
Daniel Heynderickx [4], Fabrice Cipriani [5], David Rodgers [5]

[1] Swedish Institute of Space Physics, Kiruna, Sweden. (futaana@irf.se)
[2] Max Planck Institute for Solar System Research, Germany.
[3] Kallisto Consultancy Limited, UK.
[4] DH Consultancy BVBA, Belgium.
[5] ESA, ESTEC, Netherlands.

## ABSTRACT

A new empirical, analytical model of cold plasma (< 10 keV) in the Jovian inner magnetosphere is constructed. Plasmas in this energy range impact surface charging. A new feature of this model is predicting each plasma parameter for a specified probability (percentile).

The new model was produced as follows. We start from a reference model for each plasma parameter, which was scaled to fit the data of Galileo plasma spectrometer. The scaled model was then represented as a function of radial distance, magnetic local time, and magnetic latitude, presumably describing the mean states. Then, the deviation of the observed values from the model were attribute to the variability in the environment, which was accounted for by the percentile at a given location.

The input parameters for this model are the spacecraft position and the percentile. The model is inteded to be used for the JUICE mission analysis.

## 1. Objective

The objective of this study is to create an empirical cold plasma environment model of the Jovian magnetosphere. Using a statistical approach to the Galileo/PLS dataset, the cold plasma parameters are expressed in analytical forms. The formulation also offers probability model in order to assess the environment under extreme conditions. Due to a strong interest of application to the JUpiter ICy moons Explorer (JUICE) mission, the study focuses on the distance range between 9 Rj (Europa orbit) and 30 Rj.

## 2. Dataset

Galileo PLS data [1] was used. For ion data analysis, pre-processed parameter dataset [2] was used. For electron data analysis, we reproduce the spectrum following the instruction in PDS dataset. Then, the observed spectrum was fitted by a sum of Maxwellian and background (constant count rate over the energy) to derive the (hot) electron density and temperature (Figure 1).

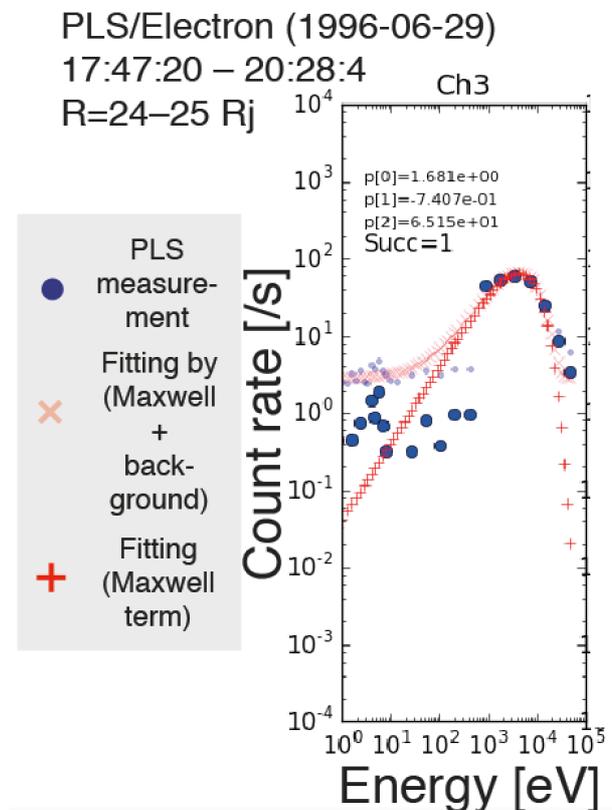

*Figure 1:* The count rate of electrons measured by PLS on June 29, 1996. The observation was done between 24–25 Rj for 2.7 hours. Translucent dots are the raw counts, while dark blue dots are the background subtracted counts. Translucent red cross the the fitting results, and the red plus (+) marks are the derived Maxwellian.

## 3. Model derivation

Each CPEM parameter was derived in three steps. First, a reference model is selected from existing models, by selecting the model best-representing the PLS data. Then, by fitting the reference model to the mean of the Galileo/PLS data, the CPEM mean model is calculated. Last, the scatter of the PLS data from the mean model is attributed to the probability of the occurrence deriving the CPEM probability model.

For example, Figure 2 shows the reference model (*Vc*,ref(*R*); taken from [3]) and the Galileo/PLS observation for the corotation flow. The data is on average represented by the reference model.

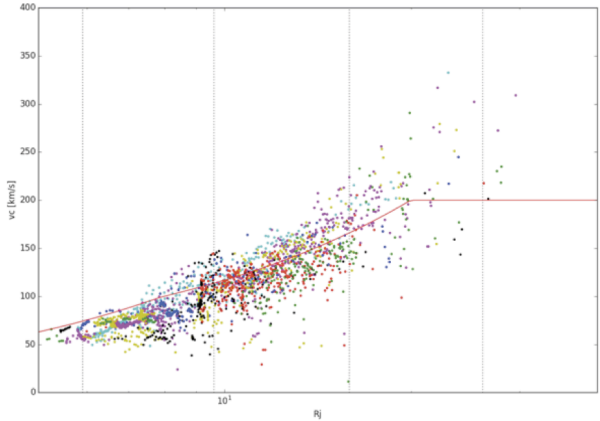

*Figure 2:* Reference model (red curve, by [3]) and PLS data for the corotational velocity model The reference model has only R-dependence. The color of the PLS data represents the MLT range of the observation (21–3 MLT for blue, 3-6 MLT for cyan, 6-9 MLT for magenda, 9-12 MLT yellow, 12-15 MLT for black, 15-18 MLT for red, and 18-21 MLT for green).

The mean model is derived by finding the best-fit parameters of the following formulation to the observed Galileo/PLS data.

$$\frac{Vc,mean(R,MLT,MLAT)}{Vc,ref(R)} = \alpha_0 +$$

$$\alpha_1 \sin(MLT - \alpha_2) +$$

$$R(\alpha_3 + \alpha_4 \sin(MLT - \alpha_5))$$

Here *Vc*,mean is the mean model that we obtained. *Vc*,ref is the reference model for the convection flow (in this case taken from [3]). *R*, *MLT*, and *MLAT* is the distance, magnetic local time, and magnetic latitude of the spacecraft position. $\alpha_0$ to $\alpha_5$ is the parameters that we obtained in this study.

Then, the Galileo/PLS data is again normalized by the mean model to obtain the derivation for each data. The deviation are used to obtain the probability model (*Vc*,prob; Figure 3).

$$Vc, prob(R, MLT, MLAT, p) = (\alpha_6 + p \cdot \alpha_7) \times$$

$$Vc, mean(R, MLT, MLAT)$$

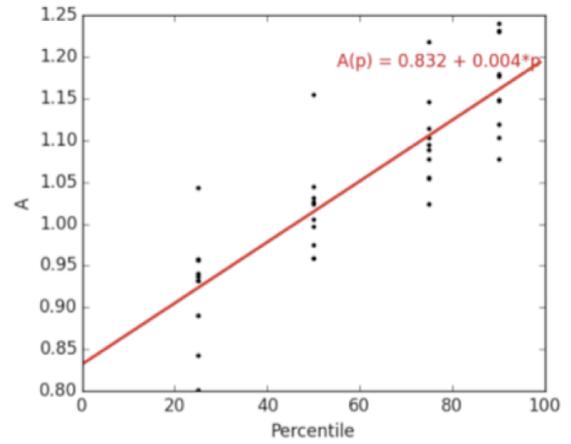

*Figure 3:* The percentile distribution of the radial velocity normalized by the mean model. The red straight line provides the percentile correction coefficient.

## 4. Results

Similarly, we derived the formulations for other cold plasma parameters. Figure 4 shows a sample plot of the mean model (probability 50%) of the density in the plane of MLT=0h.

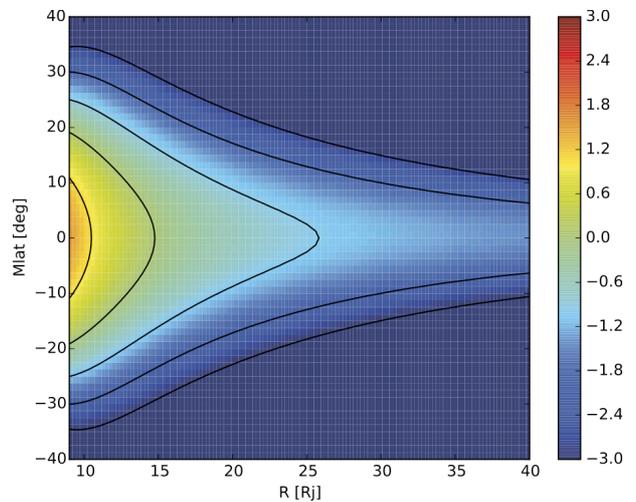

*Figure 4:* Median model in 2D representation at MLT=0h.

## 5. Concluding remarks

We present an approach to derive the new empirical probability model of the cold plasma environment of the Jovian magnetosphere with a main focus on the way of derivation of the model. Summarized concept of the CPEM is shown in Figure 5. From Galileo/PLS data we could derive a part of cold plasma parameters in analytic form. On the other hand, there still remain some parameters that could not be addressed. For example, deriving cold (<100 eV) electron moments are not straightforward. All the possible instrumental effects (e.g. background, degradation, and spacecraft potential) should be removed spectrum-by-spectrum in the PLS dataset. Thus, cold electron models (<100 eV) have not been formulated. Otherwise, we have new empirical probability model. The full CPEM model is going to be released with the SPENVIS integration. Contact to Dr Futaana (futaana@irf.se) for up-to-date information.

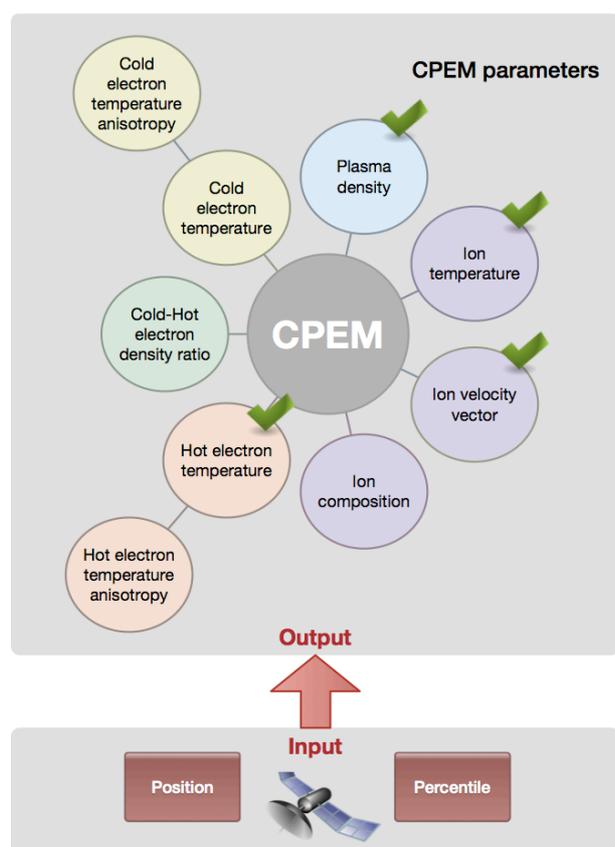

**Figure 5:** *The concept of CPEM probability (percentile) model. The input is the spacecraft position and the percentile. From Galileo data analysis, we obtained an empirical, analytic models for plasma density, ion temperature, ion velocity vector (each component independently), and hot (~keV) electron temperature.*

## 6. REFERENCES

1. Frank, L. A., K. L. Ackerson, J. A. Lee, M. R. English, and G. L. Pickett, The plasma instrumentation for the Galileo mission, *Space Sci. Rev.*, 60, 283–304, doi:10.1007/BF00216858, 1992.

2. Bagenal, F., and P. A. Delamere, Flow of mass and energy in the magnetospheres of Jupiter and Saturn, *J. Geophys. Res,* 116, A05209, doi:10.1029/2010JA016294, 2011.

3. Divine, N., and H. B. Garrett, Charged particle distributions in Jupiter's magnetosphere, *J. Geophys. Res.*, 88, 6889–6903, doi:10.1029/JA088iA09p06889, 1983.

## 7. Acknowledgement

Acknowledgement: This work has been conducted under the European Space Agency program, Juice Charging Analysis Tool (JCAT), lead by Dr David Rodgers, Dr Fabrice Cipriani, and Dr Nathelie Burdin of ESA-ESTEC (ESA Contract 4000109999/13/NL/AK). We greatly appreciate to the Galileo/PLS group, with particular emphasis of the Principal Investigator, Prof L. A. Frank. The PLS data was obtained from the Planetary Data System (PDS). Last but not least we acknowledge Prof Fran Bagenal and her colleagues to the provision of processed PLS data, which were also used for a part of this study.